\documentclass[final,5p]{elsarticle}
\usepackage{amsmath,amssymb,bm}
\usepackage{graphicx,color,ulem}

\journal{Physica E}

\begin{document}

\begin{frontmatter}

\author{Ai Yamakage \corref{cor1}}
 \ead{ai@rover.nuap.nagoya-u.ac.jp}
\cortext[cor1]{Corresponding author. Address: Department of Applied Physics, Nagoya University, Nagoya 464-8603, Japan. Tel.: +81 52 798 3701; fax: +81 52 789 3298.}

\title{Interference of Majorana fermions in NS junctions}


\author{Masatoshi Sato}

\address{Department of Applied Physics, Nagoya University, Nagoya 464-8603, Japan}

\begin{abstract}
We investigate interference of Majorana fermions (MFs) in NS junctions.
A general formula of charge conductance $G$ for NS junctions with two MFs
is derived based on the low energy effective model.
It is  found that $G$ for two MFs takes various values $0 \leq G \leq
4e^2/h$ owing to interference of the MFs, while $G$ is quantized as
$G=2e^2/h$ for a single MF. 
The value of $G$ is determined by symmetry of the system.
%
%
As an example, we investigate the complete destructive interference of
two degenerate MFs reported by Ii et al. [A. Ii, A. Yamakage, K. Yada, M. Sato, and
Y. Tanaka, Phys. Rev. B \textbf{86}, 174512 (2012)], and identify the symmetry
responsible for the destructive interference. 
\end{abstract}


\end{frontmatter}

\section{Introduction}

Majorana fermion (MF) is a particle long thought to exist since
the prediction by Ettore Majorana \cite{majorana37}. 
In spite of many efforts, MFs have not been
identified yet as an elementary particle. 
On the other hand, in recent years, topological superconductors (TSCs)
with a non-trivial bulk
topological invariant have been found
to host MFs as a collective excitation
\cite{wilczek09,franz10,qi11,alicea12,tanaka12,beenakker13}. 
In particular, spin--orbit coupled $s$-wave TSCs have attracted much attention in
this context \cite{sato03, fu08, sato09, sato10, sau10, alicea10,
lutchyn10, oreg10, mourik12, deng12,das12}. 
One of the interesting features of MFs is the non-Abelian anyon statistics
\cite{ivanov01}, which  can be applied to the fault tolerant topological quantum
computation \cite{kitaev06, nayak08}.

To identify MFs in TSCs, NS junctions have been considered.
Indeed, if an NS junction supports a single MF, the charge conductance
through the NS
junction shows a distinct zero-bias
conductance peak \cite{tanaka95, bolech07, nilsson08, law09, benjamin10, sasaki11,  yamakage12}. 
However, topological analyses
\cite{schnyder08,kitaev09,schnyder09,ryu10} have indicated that multiple
MFs can show up in various topological phases if the systems are
characterized by an integer topological number $\mathcal N \in \mathbb Z \geq 2$.
Then  a natural question is ``how to identify
multiple MFs?". 
One might expect that $\mathcal N$ MFs show a zero-bias
conductance peak that is $\mathcal N$ times larger than that for a
single MF, but this is not always the case.
It has been reported that the zero-bias conductance in NS junctions
with two MFs can vanish for a quantum anomalous Hall system \cite{ii12} and for an effective model with single channel in the normal metal \cite{ioselevich13}. 
Therefore, more detailed analysis on the charge transport in a  system
with two MFs is needed.

In this paper, we reveal that two MFs show a variety of
the tunneling conductance owing to the destructive interference. 
The conductance is determined by symmetry of the system.
This result implies that symmetry consideration on MFs is essential
to understand transport experiments of NS junctions with multiple
MFs.

The paper is organized as follows.
First, we introduce the effective model and derive the charge
conductance of an NS junction with two MFs in Sec.\ref{sec2}.
Next, in Sec.\ref{sec3}, we review the NS junction of a superconducting
quantum anomalous Hall insulator and apply our effective theory to it.
The conductance vanishing is explained in the viewpoint of symmetry of the MFs.
Finally,  we summarize our results and discuss a future perspective in Sec.\ref{summary}.

\section{Charge conductance for two Majorana fermions}
\label{sec2}

Here we derive a general formula for the charge conductance of the
effective model describing a NS junction with two Majorana fermions.
The behavior of the charge conductance is discussed based on this 
effective theory.

\subsection{Effective model of the NS junction}

We extend the effective model \cite{bolech07, nilsson08, law09, ioselevich13} of NSN junctions to that of a NS junction with two MFs as an Andreev bound state. 
In the low energy, 
quasiparticles in the bulk superconductor are neglected.
The Hamiltonian reads
\begin{align}
 H &= \sum_{k \sigma} \xi_k c^\dag_{k \sigma} c_{k \sigma}
     + i E_{\rm M} \gamma_1 \gamma_2 - i E_{\rm M} \gamma_2 \gamma_1
 \nonumber\\ & \quad
 + \sum_{k \sigma}
     \sum_{i=1}^2
     \left(
      t_{\sigma i} c^\dag_{k \sigma} - t_{\sigma i}^* c_{-k \sigma}
     \right) \gamma_i,
 \label{hmlt}
\end{align}
where $c_{k \sigma}$ is an annihilation operator of an electron in the
normal metal with momentum $k$ and spin $\sigma = \uparrow, \downarrow$,
and $\gamma_i$ $(i=1,2)$ denote MFs satisfying $\gamma_i=\gamma_i^{\dagger}$.
The first term of the above equation is the kinetic energy of the normal
metal, and  
$\xi_k$ is the kinetic energy measured from the Fermi level $\mu$.
The second and third terms denote the hybridization between the MFs.
$E_{\rm M} \in \mathbb R$ denotes the strength of the hybridization.
The last term represents the hybridization between the electron (or hole)
and the MFs. Here the strength $t_{\sigma i}$ depends on the connectivity of
interface of the NS junction, which can be experimentally controlled. 
Also, the momentum dependence of $t_{\sigma
i}$ is neglected since the MFs are localized at surface of the
superconductor. 
This model corresponds to the Anderson model with particle-hole
symmetric localized states. 
Thus the standard method for the Anderson model can be applied to the
present system, as explained in the following section.

Let us mention that the from of the effective Hamiltonian is severely
restricted by the Majorana condition $\gamma_i^\dag = \gamma_i$.
This condition determines the form of the hybridization between MFs
as $i E_{\rm M} \gamma_1 \gamma_2 - i E_{\rm M} \gamma_2 \gamma_1$ with
real $E_{\rm M}$, and 
the relation $(t_{\sigma i} c^\dag_{k \sigma} \gamma_i)^\dag =
t^*_{\sigma i} \gamma_i c_{k \sigma}$ leads the last
term of Eq. (\ref{hmlt}).

\subsection{Charge conductance}
\label{sec:Cc}

Now we calculate the charge conductance using the effective model.
The last term of Eq. (\ref{hmlt}) yields that  an electron (hole) is
converted to a MF with the transition amplitude $t_{\sigma i}$
($-t^*_{\sigma i}$), and  
the inverse process is also possible. 
This means that an electron can transit to an hole mediated by a MF.
This process, i.e., the Andreev reflection contributes to charge
transport with charge flow $2e$. 
The corresponding conductance is given by \cite{nilsson08}  
\begin{align}
 G(\omega) = \frac{2e^2}{h} \sum_{\sigma \sigma'} |T_{\sigma \mathrm e,\sigma' \mathrm{h}}|^2,
\end{align}
with
\begin{align}
\hat T&=
\begin{pmatrix}
 T_{\uparrow \mathrm e, \uparrow \mathrm e} & T_{\uparrow \mathrm e, \downarrow \mathrm e} & T_{\uparrow \mathrm e, \uparrow \mathrm h} & T_{\uparrow \mathrm e, \downarrow \mathrm h}
 \\
T_{\downarrow \mathrm e, \uparrow \mathrm e} & 
T_{\downarrow \mathrm e, \downarrow \mathrm e} & 
T_{\downarrow \mathrm e, \uparrow \mathrm h} & 
T_{\downarrow \mathrm e, \downarrow \mathrm h}
\\
T_{\uparrow \mathrm h, \uparrow \mathrm e} & 
T_{\uparrow \mathrm h, \downarrow \mathrm e} & 
T_{\uparrow \mathrm h, \uparrow \mathrm h} & 
T_{\uparrow \mathrm h, \downarrow \mathrm h}
\\
T_{\downarrow \mathrm h, \uparrow \mathrm e} & 
T_{\downarrow \mathrm h, \downarrow \mathrm e} & 
T_{\downarrow \mathrm h, \uparrow \mathrm h} & 
T_{\downarrow \mathrm h, \downarrow \mathrm h}
\end{pmatrix}
\nonumber\\
&=
\hat w
\frac{2}{\omega- \hat H_{\rm M}+i \hat w^\dag \hat w} \hat w^\dag.
\label{tmat}
\end{align}
Here $\omega = eV$ with the bias voltage $V$, and 
the matrices $\hat w$ and $\hat H_{\rm M}$ are defined by
\begin{align}
\hat w = \begin{pmatrix}
  \tilde t_{\uparrow 1} & \tilde t_{\uparrow 2}
  \\
  \tilde t_{\downarrow 1} & \tilde t_{\downarrow 2}
  \\
  -\tilde t_{\uparrow 1}^* & - \tilde t_{\uparrow 2}^*
  \\
  - \tilde t_{\downarrow 1}^* & - \tilde t_{\downarrow 2}^*
 \end{pmatrix},
\end{align}
and
\begin{align}
 \hat H_{\rm M}
 = \begin{pmatrix}
  0 & iE_{\rm M}
  \\
  -i E_{\rm M} & 0
 \end{pmatrix},
\end{align}
respectively.
$\tilde t_{\sigma i}$ is defined by $\tilde t_{\sigma i} = \sqrt{\pi
\rho_0} t_{\sigma i}$ with $\rho_0$ being the density of states at the
Fermi level.
The explicit form of $G(\omega)$ is easily obtained. (See. \ref{derivation}).

First, let us mention the conductance for a $s_z$-conserving NS junction, which is realized in mirror symmetric TSCs \cite{ueno13, chiu13, zhang13}.
In the $s_z$-conserving system, one of the MFs has spin up while the
other has spin down, and the system is decoupled into the up spin and
the down spin sectors. 
 Consequently, one obtains 
$t_{\uparrow 1}, t_{\downarrow 2} \ne 0, t_{\downarrow 1} = t_{\uparrow 2} = 0$.
The resulting zero--bias conductance is always given by $G = 4 e^2/h$.

In the following, we focus on several important cases with $\omega=0$
and $E_{\rm M}=0$. 
In the below,  $\theta_i$ is the phase of the hybridization amplitude;
$t_{\uparrow i} = t_i e^{i \theta_i}$.

\noindent 
\begin{description}
\item[(I)]  Unitary case: $ t_{\uparrow i} = \eta_i t_{\downarrow i}$.
\begin{description}
 \item[a)] same sign case: $\eta_1 \eta_2 = 1$. 
\begin{align}
T_{\sigma \mathrm e, \sigma' \mathrm h}
= 
\begin{cases}
\displaystyle\frac{i}{2} e^{i2\theta_1}, &
(\theta_1 - \theta_2)/\pi \in \mathbb Z,
\\
0, & (\theta_1 - \theta_2)/\pi \notin \mathbb Z,
\end{cases}
\end{align}

\item[b)] different sign case: $\eta_1 \eta_2 = -1$.
\begin{align}
 T_{\uparrow \mathrm e, \uparrow \mathrm h}
 &=
 T_{\downarrow \mathrm e, \downarrow \mathrm h}
 = i e^{i(\theta_1+\theta_2)} \cos(\theta_1-\theta_2),
 \\
 T_{\uparrow \mathrm e, \downarrow \mathrm h}
 &=
 T_{\downarrow \mathrm e, \uparrow \mathrm h}
 = \eta_2 e^{i(\theta_1+\theta_2)} \sin(\theta_1-\theta_2).
\end{align}

\end{description}

\item[(II)]Anti-unitary case:  $t_{\uparrow i} = \eta_i
	   t_{\downarrow i}^*$.

\begin{description}
 \item[a)] same sign case: $\eta_1 \eta_2 = 1$.
\begin{align}
 T_{\uparrow \mathrm e, \uparrow \mathrm h}
 &= -T_{\downarrow \mathrm e, \downarrow \mathrm h}^*
 = \begin{cases}
\displaystyle\frac{i}{2} e^{i2\theta_1}, 
& (\theta_1-\theta_2)/\pi   \in \mathbb Z,
   \\
   0, & (\theta_1-\theta_2)/\pi   \notin \mathbb Z,
 \end{cases}
 \\
 T_{\uparrow \mathrm e, \downarrow \mathrm h}
 &=
 T_{\downarrow \mathrm e, \uparrow \mathrm h}
 = \begin{cases}
   \displaystyle\frac{i \eta_1 }{2}, 
& (\theta_1-\theta_2)/\pi   \in \mathbb Z,
   \\
   i\eta_1, 
& (\theta_1-\theta_2)/\pi   \notin \mathbb Z.
 \end{cases}
\end{align}

\item[b)] different sign case: $\eta_1 \eta_2 = -1$.
\begin{align}
 T_{\uparrow \mathrm e, \uparrow \mathrm h}
 &= -T_{\downarrow \mathrm e, \downarrow \mathrm h}
 = i e^{i(\theta_1+\theta_2)} \cos(\theta_1-\theta_2),
 \label{t1}
 \\
 T_{\uparrow \mathrm e, \downarrow \mathrm h}
 &= 
 T_{\downarrow \mathrm e, \uparrow \mathrm h} = 0.
 \label{t2}
\end{align}

\end{description}

\end{description}

\begin{table}
\centering
\begin{tabular}{c|cc}
 \hline\hline
& Unitary & Anti-unitary
 \\
$\eta_1 \eta_2$ & $t_{\uparrow i} = \eta_i t_{\downarrow i}$ & $t_{\uparrow i} = \eta_i t_{\downarrow i}^*$
  \\ 
  \hline
$+1$ & $0$ & $4e^2/h$
  \\
$-1$ & $4e^2/h$ & $2\cos^2(\theta_1-\theta_2) 2e^2/h$
 \\
 \hline\hline
\end{tabular}
\caption{Zero-bias charge conductance for NS junctions with two Majorana fermions. $\theta_i$ is the phase of the hybridization amplitude;  $t_{\uparrow i} = t_i e^{i \theta_i}$.
Here, we assume that $(\theta_1 - \theta_2)/\pi \notin \mathbb Z$ for
 $\eta_1 \eta_2 = +1$ in both the unitary and anti-unitary cases.
}
\label{table1}
\end{table}

The resulting conductance is summarized in Table \ref{table1}.
It has been known that a system supporting only a single MF shows the
universal conductance $2e^2/h$, due to the resonant Andreev reflection
\cite{kashiwaya00, bolech07, nilsson08, law09}. 
In contrast, 
in the case of two MFs, the resonance results in  various values of the
conductance: 
i) $4 e^2/h$, ii) $2 e^2/h$, iii) 0, and iv) $4
\cos^2(\theta_1-\theta_2) e^2/h$.
The first one is constructive interference, while others are destructive
interference between MFs. 
The last case ${\bf (II)}$ ${\bf b)}$, where the anti-unitary condition ($t_{\uparrow i} =
\eta_i t^*_{\downarrow i}$) and $\eta_1 \eta_2 = -1$ are satisfied, is
special.
In this case, the resulting
conductance is determined only by the phase difference
$\theta_1-\theta_2$, due to interference of MFs. 
If $(\theta_1-\theta_2)/\pi \in \mathbb Z+1/2$, then the conductance
completely vanishes. 
This peculiar behavior is robast against the hybridization between the MFs ($E_{\rm M} \ne 0$).
When $E_{\rm M} \ne 0$, Eqs. (\ref{t1}) and (\ref{t2}) are modified as
\begin{align}
 T_{\uparrow \mathrm e, \uparrow \mathrm h}
 &= -T_{\downarrow \mathrm e, \downarrow \mathrm h}
 = i \frac{e^{i(\theta_1+\theta_2)}}{1+ \kappa^2}  \cos(\theta_1-\theta_2),
 \\
 T_{\uparrow \mathrm e, \downarrow \mathrm h}
 &= 
 T_{\downarrow \mathrm e, \uparrow \mathrm h} = -i
 \frac{\kappa}{1+\kappa^2} \cos(\theta_1-\theta_2). 
\end{align}
with $\kappa = E_{\rm M}/(4 \tilde t_1 \tilde t_2)$, but 
the resulting conductance remains the same, i.e., $4
\cos^2(\theta_1-\theta_2) e^2/h$.
Therefore, the conductance vanishes again when $(\theta_1-\theta_2)/\pi \in
\mathbb Z+1/2$.
We will see that the last case is realized in a superconducting quantum
anomalous Hall insulator in the next Section.

\section{Conductance vanishing in a superconducting quantum anomalous Hall insulator}
\label{sec3}

In the previous section, we present general results of the charge
conductance for NS junctions with two MFs. 
The conductance depends the phase differences of the couplings
between the electron in the normal metal and the MFs on the interface.
The phase differences are restricted by symmetry of
the system. 
Here, we apply the results to the superconducting quantum anomalous Hall
(QAH) insulator, in which the destructive interference occurs.

\subsection{Model}

Based on the Hamiltonian of a QAH insulator, 
\begin{align}
 H_{\rm QAH}(k_x,k_y) &= m(k_x,k_y) \sigma_z + A (k_x \sigma_x + k_y \sigma_y), 
\end{align}
the Hamiltonian of a superconducting QAH insulator \cite{qi10} is given by
\begin{align}
&H_{\rm SQAH}(k_x, k_y) 
\nonumber\\
=&
 \begin{pmatrix}
  H_{\rm QAH}(k_x,k_y) - \mu  & i \sigma_y \Delta
  \\
  -i \sigma_y \Delta & -H_{\rm QAH}^*(-k_x,-k_y) + \mu
 \end{pmatrix}
 \nonumber\\
= &
 -\mu \tau_z + m(k_x ,k_y) \sigma_z \tau_z + A(k_x \sigma_x + k_y \sigma_y \tau_z) - \Delta \sigma_y \tau_y,
\end{align}
with $m(k_x,k_y) = m_0 + B (k_x^2+k_y^2)$. Here 
$m_0$ is the band gap between the conduction and the valence bands, 
$1/2B$ is the effective masses of these bands, which
are the same in this model, for simplicity,  
$\mu$ is the chemical potential controlled by the carrier doping,
and $\Delta$ is the pair potential induced by an $s$-wave
superconductor attached to the QAH insulator.
$\tau_i$ and $\sigma_i$ are  the Pauli matrices in the Nambu space and the
spin space, respectively.
Hereafter we assume $B>0$ without loss of generality.

The topological phase of this system is characterized by the Chern
numbers \cite{qi10}. 
For the QAH system $H_{\rm QAH}$,
the Chern number $N$ is defined by
\begin{align}
 N = \frac{1}{2\pi}
 \int d^2k \left(
  \frac{\partial a_y(\bm k)}{\partial k_x}
  -
  \frac{\partial a_x(\bm k)}{\partial k_y}
 \right),
\end{align}
where $a_i(\bm k) = -i \langle \bm k | \partial/\partial k_i | \bm k
\rangle$ and $|\bm k \rangle$ is the eigenvector of $H_{\rm QAH}(\bm k)$
for the valence band.
The Chern number is given by
$N=0$ for $m_0>0$ (nontopological insulator) and $ N = 1$ (QAH
insulator) for $m_0<0$ (Table \ref{pd_qah}).
In a similar manner, the Chern number $\mathcal N$ for the
superconducting state $H_{\rm SQAH}$ is defined. 
Note that when $\Delta=0$, $\mathcal N$ reduces to $2N$ since the hole
component in the BdG Hamiltonian redundantly contributes to the Chern
number. 
In the case of $\Delta \ne 0$, the Chern number $\mathcal N$ is obtained to be
 $\mathcal N = 0$  for $m_0 > \sqrt{\Delta^2+\mu^2}$, $\mathcal N = 1$
 for $|m_0| < \sqrt{\Delta^2 + \mu^2}$, and $\mathcal N = 2$ for $m_0 <
 -\sqrt{\Delta^2 + \mu^2}$ (Table \ref{pd_sqah}).
From the bulk-edge correspondence, there are $\mathcal N$ Majorana edge
states in each phase.
\begin{table}
\centering
\begin{tabular}{c|c}
\hline\hline
 Mass term & Chern number
\\ \hline
 $m_0 > 0$ & $N = 0$
 \\
 $m_0 < 0$ & $N = 1$
\\ \hline \hline
\end{tabular}
\caption{Topological phase of the normal state $H_{\rm QAH}$.}
\label{pd_qah}
\end{table}
\begin{table}
\centering
\begin{tabular}{c|c}
  \hline\hline
  Mass term &  Chern number
  \\
  \hline
  $m_0 > +\sqrt{\Delta^2 + \mu^2}$ & $\mathcal N = 0$
  \\
  \hspace{-1.2ex}$|m_0| < +\sqrt{\Delta^2 + \mu^2}$ & $\mathcal N = 1$
  \\
  $m_0 < - \sqrt{\Delta^2 + \mu^2}$ & $\mathcal N = 2$
  \\
  \hline\hline
\end{tabular}
\caption{Topological phase of the superconducting state $H_{\rm SQAH}$.}
\label{pd_sqah}
\end{table}

\begin{figure*}
\centering
\includegraphics{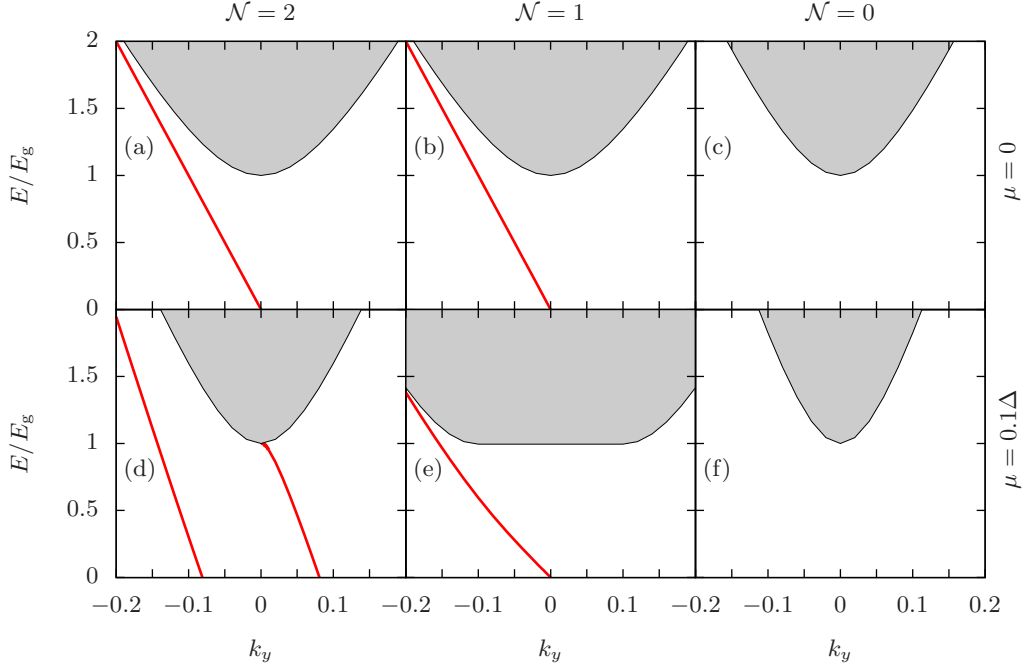}
\caption{Energy dispersions for the $\mathcal N = 2$ (a,d), $\mathcal N = 1$ (b,e), and $\mathcal N = 0$ (c,f) phases. 
The solid lines and shaded region denote those for the MFs at the edge and 
 for the bulk states, respectively.
The chemical potential is set to $\mu = 0$ (a,b,c) and $\mu=0.1\Delta$ (d,e,f).
The mass term is set to $m_0 = -0.2$ for $\mathcal N=2$, $m_0 = 0$ for $\mathcal N=1$, and $m_0 = 0.2$ for $\mathcal N=0$.
The other parameters are taken as follows.
$A=B=1$,   $\Delta=0.1$.
}
\label{energy}
\end{figure*}

The bulk and edge energy dispersions are shown in Figs.\ref{energy}(a)-(f) for
the $\mathcal N = 2$, 1, and 0 phases. 
For $\mu=0$, while the systems with $\mathcal N=2$ and $\mathcal N =1$
(Figs. \ref{energy}(a) and \ref{energy}b) have the similar energy
dispersions, two MFs are degenerated in the momentum space in the case
of $\mathcal N=2$. 
On the other hand, for $\mu \ne 0$, the degeneracy of the MFs in the case of
${\mathcal N}=2$ is lifted
(Fig. \ref{energy}d) since 
the chemical potential gives rise to an addtional coupling between the MFs.

\subsection{Mapping to the effective model}

The electronic and transport properties of the bulk
\cite{qi10,ii11,ii12} and the edge \cite{chung11} in the superconducting
QAH insulator have been extensively studied.
Below, we discuss why the conductance vanishes \cite{ii12} in the
$\mathcal N=2$ phase from the viewpoint of the effective theory
discussed in the previous section.
In the low transmissivity limit of the NS junction, the Andreev
reflection occurs only through MFs localized at the interface.
The NS junction can be mapped to the effective model in this limit.
For instance, such a situation is realized for NS junctions with a thick oxide layer between the normal metal and superconductor.

For this mapping,  let us first examine the symmetry of the
QAH insulator and its superconducting state. 
Because of an internal magnetization given by $m(k_x,k_y)\sigma_z$,
neither the
time-reversal symmetry or the two-fold rotational symmetry with respect
to the $x$-axis is preserved in the QAH insulator.
Nevertheless, the combination of them is preserved, which gives a hidden
time-reversal symmetry of the system.
Correspondingly, the superconducting QAH insulator 
also has the same hidden time-reversal symmetry
since the $s$-wave gap
function does not break any symmetry. 
As a result, the BdG Hamiltonian of the superconducting QAH insulator satisfies
$\tilde \Theta H_{\rm SQAH}(k_x, k_y) \tilde \Theta^\dag = H^*_{\rm
SQAH}(-k_x, k_y)$ with $\tilde \Theta = \sigma_z
\tau_z$. 
In addition, the BdG Hamiltonian has the particle-hole symmetry, 
$\tau_x H_{\rm SQAH}({\bm k}) \tau_x = -H^*_{\rm
SQAH}(-{\bm k})$ as an intrinsic symmetry of a superconductor.
Moreover, to discuss the effective model, it is convenient to consider
the hidden chiral symmetry that is obtained by combining  
the hidden time-reversal symmetry with the particle-hole symmetry.
The explicit form of the hidden chiral symmetry is given by
$\{\Gamma, H_{\rm SQAH}({\bm k})\}=0$
at $k_y=0$ with $\Gamma = \tilde\Theta
\tau_x = \sigma_z \tau_y$.
We dub the eigenvalue of $\Gamma=\pm 1$ as chirality.

When the system has such a chiral symmetry, the BdG Hamiltonian becomes
off-diagonal if we take the basis where $\Gamma$ is diagonal.  
This restricts possible couplings of the system.
Indeed, only couplings between states with opposite chiralities are
possible. 

Since the effective model of NS junctions we considered is
one-dimensinal, we perform the dimensional reduction of the superconducting QAH
insulator by fixing $k_y$ as $k_y=0$.
The resultant one-dimensional system also has the hidden chiral
symmetry.
It also has the  quasiparticle spectrum
of the superconducting QAH insulator at $k_y=0$.
Therefore, as illustrated in Fig.\ref{energy}(a), 
there are two degenerate
Majorana zero modes $\gamma_1$ and $\gamma_2$ in the ${\cal N}=2$ phase
when $\mu=0$.

When $\mu\neq 0$, the two Majorana zero modes are gapped, as is seen in
Fig.\ref{energy} (d).
Therefore, the chemical potential $\mu$ induces the mass term, $iE_{\rm
M}\gamma_1\gamma_2-iE_{\rm M}\gamma_2\gamma_1 $, in
Eq.(\ref{hmlt}). 
As was mentioned in the above, since the hidden chiral symmetry admits only
the coupling between states with opposite chiralities,  this means that
$\gamma_1$ and $\gamma_2$ have opposite chiralities.
Without loss of generality, we assume that $\gamma_1$ has the chirality
$\Gamma=+1$, and $\gamma_2$ has the chirality $\Gamma=-1$.

Using the hidden chiral symmetry, we can determine the couplings
between $c_{k \sigma}$ and $\gamma_i$:
In the Nambu basis, $(c_{k \uparrow}, c_{k \downarrow}, c^\dagger_{-k
\uparrow}, c^{\dagger}_{-k \downarrow})^t$, the two independent
eigenstates with $\Gamma=1$ are given by $(1,0,i,0)^t$ and $(0,1,0, -i)^t$, which correspond to the
operators, $c_{k \uparrow}-ic^{\dagger}_{-k\uparrow}$ and $c_{k
\downarrow}+ic_{-k\downarrow}^{\dagger}$, respectively, and  
the eigenstates with $\Gamma=-1$
are $(1,0,-i,0)^t$ and $(0,1,0, i)^t$, which correspond to 
$c_{k \uparrow}+ic^{\dagger}_{-k\uparrow}$ and $c_{k
\downarrow}-ic_{-k\downarrow}^{\dagger}$, respectively.
Thus if the hidden chiral symmetry is preserved, the possible couplings are
$(c_{k \uparrow}+ic^{\dagger}_{-k\uparrow})\gamma_1$, $(c_{k
\downarrow}-ic_{-k\downarrow}^{\dagger})\gamma_1$, $(c_{k
\uparrow}-ic^{\dagger}_{-k\uparrow})\gamma_2$ and $(c_{k
\downarrow}+ic_{-k\downarrow}^{\dagger})\gamma_2$.
In terms of $t_{\sigma i}$ in Eq.(\ref{hmlt}), 
these couplings imply that
\begin{align}
t_{\uparrow 1}=-it^*_{\uparrow 1},
\quad
t_{\downarrow 1}=it^*_{\downarrow 1},
\\
t_{\uparrow 2}=it^*_{\uparrow 2},
\quad
t_{\downarrow 2}=-it^*_{\downarrow 2}.
\label{eq:chiralrelation}
\end{align}
From these relations, we obtain
\begin{align}
G=G^{(\pm)}&\equiv \frac{4e^2}{h}\left[1-\frac{(|t_{\uparrow
				   1}||t_{\uparrow 2}| \pm|t_{\downarrow 1}||t_{\downarrow 2}|)^2}{\Gamma_1\Gamma_2}\right], 
\nonumber\\
&=\frac{4e^2}{h}
\frac{(|t_{\uparrow 1}||t_{\downarrow 2}|
\mp |t_{\uparrow 2}||t_{\downarrow 1}|)^2}{\Gamma_1\Gamma_2}
\end{align}
with $\Gamma_i=\sum_{\sigma}|t_{\sigma i}|^2$.
Here note that only $G^{(+)}$ can be zero smoothly with finite
 $t_{\sigma i}$.
This enables us to choose $G^{(+)}$ as the
conductance for the NS junction of the superconducting QAH insulator. 
Indeed, 
as is shown below, $G$ in the superconducting QAH insulator can go to zero
smoothly as $\mu\rightarrow 0$.

Let us now consider the case of $\mu=0$.
When $\mu=0$, the BdG Hamiltonian $H_{\rm SQAH}({\bm k})$ has an
accidental symmetry, $[H_{\rm SQAH}({\bm k}), \sigma_x\tau_x]=0$, while 
this symmetry is broken by the 
chemical potential $\mu$.
Combining this with hidden chiral symmetry, we can obtain another hidden
chiral symmetry, $\{H_{\rm SQHA}({\bm k}), \Gamma'\}=0$ with
$\Gamma'=\sigma_y\tau_z$, at $k_y=0$ when $\mu=0$. 
In contrast to $\Gamma$, the second chiral symmetry $\Gamma'$ is borken
by $\mu$, and thus the chemical potential term is diagonal in the basis where
$\Gamma'$ is diagonal. 
This implies that $\gamma_1$ and $\gamma_2$ have the same
chirality $\chi$ of $\Gamma'$, because the mass term $iE_{\rm
M}\gamma_1\gamma_2-iE_{\rm M}\gamma_2\gamma_1$ coresponding to the
chemical potential term $\mu$ also should be diagonal in the basis where
$\Gamma'$ is diagonal.

Here we assume that the interface between the normal metal and the
superconducting QAH insulator preserves this accidental chiral symmetry. 
Actually, in our numerical calculations in Ref.\cite{ii12} and in the next
subsection, any symmetry is not broken by the boundary condition at the interface.
Under this assumption, we can determine possible couplings between
$c_{k\sigma}$ and $\gamma_i$ in a manner similar to the $\Gamma$ case:
In the Nambu basis, $(c_{k \uparrow}, c_{k \downarrow}, c^\dagger_{-k
\uparrow}, c^{\dagger}_{-k \downarrow})^t$, the two independent
eigenstates with $\Gamma'=1$ are given by $(1,i,0,0)^t$ and $(0,0,1,
-i)^t$, which correspond to the operators, $c_{k
\uparrow}-ic_{k\downarrow}$ and $c_{-k
\uparrow}^{\dagger}+ic_{-k\downarrow}^{\dagger}$, respectively, and 
the eigenstates with $\Gamma'=-1$
are $(1,-i,0,0)^t$ and $(0,0,1, i)^t$, which correspond to 
$c_{k \uparrow}+ic_{k\downarrow}$ and $c_{-k
\uparrow}^{\dagger}-ic_{-k\downarrow}^{\dagger}$, respectively.
Therefore, the possible couplings are
$(c_{k \uparrow} + \chi ic_{k\downarrow})\gamma_1$, 
$(c_{k \uparrow} + \chi ic_{k\downarrow})\gamma_2$, 
$(c_{-k \uparrow}^{\dagger} - \chi ic_{-k\downarrow}^{\dagger})\gamma_1$, 
and
$(c_{-k \uparrow}^{\dagger} - \chi ic_{-k\downarrow}^{\dagger})\gamma_2$.
In terms of $t_{\sigma i}$ in Eq.(\ref{hmlt}), 
these couplings give
\begin{align}
t_{\uparrow 1} = i\chi t_{\downarrow 1},
\quad
t_{\uparrow 2} = i\chi t_{\downarrow 2}.
\label{eq:chiralrelation2}
\end{align}

From Eqs. (\ref{eq:chiralrelation}) and (\ref{eq:chiralrelation2}),
we can see that $G=0$ when $\mu=0$ in the effective model of the
superconducting QAH insulator:
These relations yield 
\begin{align}
t_{\uparrow 1}= -\chi t_{\downarrow 1}^{*}, 
\quad
t_{\uparrow 2}= \chi t_{\downarrow 2}^{*}, 
\quad 
\theta_1-\theta_2=\frac{\pi}{2}+n\pi,
\end{align}
with $n\in \mathbb Z$, and thus
the present case reduces to the anti-unitary case with $\eta_1\eta_2=-1$
in Sec.\ref{sec:Cc}. 
Therefore, the conduction, $G=(4 e^2/h)\cos^2 (\theta_1-\theta_2)$,
vanishes. 

Before closing this subsection, we would like to make some comments.
(1) First, the above discussions reveal that the complete destructive
interference of MFs in the  ${\cal N}=2$ phase is not robust. While 
the accidental symmetry is needed to obtain $G=0$,  this symmetry is
easily broken by the chemical potential $\mu$.
To make things worse,  the normal metal attached to the NS junction does
not have such an accidental symmetry. 
Thus an ignored interaction in a real system also may break the
accidental symmetry. 
(2) In spite that the complete destructive interference cannot be
expected in a real system, as we mentioned above, the destructive
interference of MFs can be expected in
the ${\cal N}=2$ phase of the superconducting QAH insulator. 
Since the hidden chiral symmetry $\Gamma$ is originated from the
remnant of the two fold-rotation symmetry, it is not broken as far as
the NS junction respects the two-fold rotation symmetry.
In this sense, $\Gamma$ is intrinsic. 
Because the conductance $G=G^{(+)}$ is less than $4e^2/h$, 
our effective model predicts that the interference of two MFs is always
destructive in the ${\cal N}=2$ phase of the superconducting QAH
insulator.



\subsection{Comparison between the original and effective models}

\begin{figure}
\centering
\includegraphics[scale=0.2]{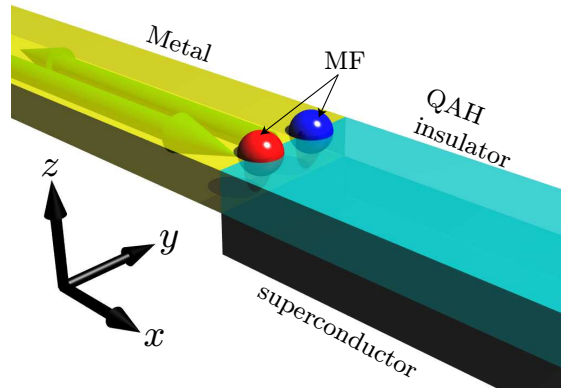}
\caption{Schematic of the NS junction for $\mathcal N=2$.
Charge current along the $x$-axis is prohibited due to interference of MFs.
}
\label{ns}
\end{figure}
\begin{figure*}
\centering
\includegraphics{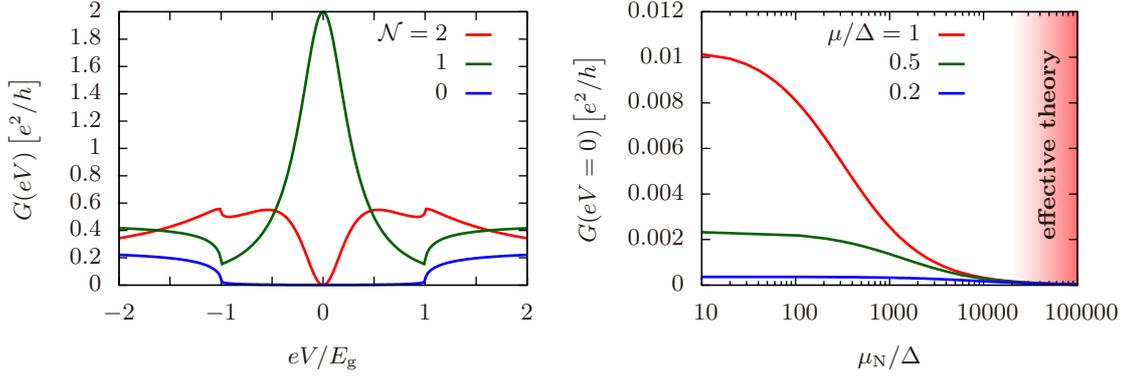}
\caption{Charge conductance $G$
in unit of $e^2/h$ for a one-dimensional NS junction of SQAH insulator for the $\mathcal N = 2$, 1, and 0 phases. 
The left panel shows $G$ as a function of the bias voltage $V$ for $\mu=0$ and $\mu_{\rm N}=100$. 
The right panel shows $G$ as a function of the chemical potential $\mu_{\rm N}$ of the normal metal for $V=0$ and $\mathcal N = 2$.
$E_{\rm g}$ is the energy gap of the SQAH insulator and $\Delta$ is the pair potential.
The mass term is set to $m_0 = -0.2$ for $\mathcal N=2$, $m_0 = 0$ for $\mathcal N=1$, and $m_0 = 0.2$ for $\mathcal N=0$.
The other parameters are taken as follows.
$A=B=1$,   $\Delta=0.1$, $m_{\rm N} B = 1$.
In $\mu_{\rm N} \to \infty$ limit, which corresponds to the low transmissivity limit, the effective theory proposed in the paper becomes exact.
}
\label{zbcd}
\end{figure*}

To confirm the validity of the effective model, we compare the conductance obtained from the effective
model and from those form the original model described by $H_{\rm SQAH}$.
The NS junction is illustrated in Fig. \ref{ns}.
For comparison, we consider the one-dimensional NS junction ($k_y=0$) as
in the previous subsection.
The Hamiltonian of the NS junction is given by
\begin{align}
H(x<0) &= \left(\frac{k_x^2}{2m_{\rm N}} - \mu_{\rm N} \right) \tau_z,
\label{ns1}
\\
H(x>0) &= H_{\rm SQAH}(k_x \to -i\partial_x, k_y=0),
\label{ns2}
\end{align}
with $m_{\rm N}$ and $\mu_{\rm N}$ being the effective mass and the chemical
potential in the normal metal ($x<0$), respectively.
We calculate the charge conductance in the NS junction developing the
Blonder--Tinkham--Klapwijk theory \cite{ii12}. 
The obtained data are shown in Fig. \ref{zbcd}.
The left panel shows the conductance as a function of bias voltage $V$
for the $\mathcal N = 2$, 1, and 0 phases. 
The chemical potential is set to $\mu=0$.
At the zero bias voltage $V=0$, one can clearly see $G=2e^2/h$ for the
$\mathcal N=1$ phase while $G=0$ for the $\mathcal N=0$ and $\mathcal
N=2$ phases. 
Note that the conductance vanishes even though two-fold degenerated MFs
exists at the zero energy. 
The conductance vanishing at $V=0$ for $\mathcal N=2$ is consistent with
the effective theory. 

As illustrated in the right panel of
Fig. \ref{zbcd}, the zero-bias conductance in the original model takes a nonzero
value for a nonzero $\mu$ and a small $\mu_{\rm N}$. 
This behaviour is also consisitent with our consideration in the
previous section: When $\mu\neq 0$,  the relation Eq.
(\ref{eq:chiralrelation2}) does not hold since the accidental chiral
symmetry $\Gamma'$ is broken by $\mu$.
Therefore, the conductance $G$ in the effective model can be nonzero.
In addition, when $\mu_{\rm N}$ is small, the Andreev reflection can
occur directly without the mediation of MFs, because the decrease of 
$\mu_{\rm N}$ increases 
the transmissivity of the NS junction. 
This effect is beyond the range of our effective model.


\section{Discussion}
\label{summary}

In this paper we have shown that MFs can cause destructive interference, based on the effective theory focusing on the MFs.
We have derived a general formula of the conductance for NS junctions with two MFs.
In particular, for a quantum anomalous Hall system, the conductance completely vanishes when the phase
difference of the MFs is given by $\pm \pi/2$ although the density of
states of the MFs is nonzero. 
This is the direct consequence of interference of MFs.
It is worth mentioning that this interference is robust against perturbations since it stems from the chiral symmetry of the system.

One of the systems exhibiting the interference of MFs is a superconducting QAH system, which can be realized in magnetically doped/ordered topological insulators \cite{liu08, yu10, xu11, niu11, wang13}.
Recently, the QAH effect in such a system has been experimentally observed \cite{chang13}.
Our theory will be relevant to future experiments on superconducting proximity effect for QAH systems.

The present results suggest that
the conductance in systems with $N$-fold degenerated MFs is not given by
$2Ne^2/h$  (This will be discussed elsewhere). 
Generally, there are even-odd effects by the number of MFs on the conductance: if the total number of MFs is even, 
the resulting conductance at zero energy may vanish, 
but if it is odd, 
the conductance must be finite. 
This is because an unpaired MF exists in the latter case. 
However, this does not mean that the even-odd effects always occur. 
Indeed, if there are additional symmetries which stabilize MFs, 
even-odd effects can be obscure. 
For example, if the system supports time-reversal symmetry, 
the tunneling conductance cannot be zero in spite of two MFs.
A clear experimental signal for MFs is the quantized conductance $G = 2 N e^2/h$, when the $N$ MFs are somehow divided into $N$ independent sectors.
Otherwise, the MFs cause interference ($G \ne 2 N e^2/h$), thus they can form a Dirac fermion rather than MFs. 
Then a careful analysis, e.g., symmetry consideration, is needed to
correctly understand the zero-bias conductance for multiple MFs
systems.

\textit{Note added}.
Upon completing the manuscript, we became aware of a work that discuss the complete destructive interference of MFs in the SQAH insulator
 by J. J. He, J. Wu, T. P. Choy, X.-J. Liu, Y. Tanaka, and K. T. Law \cite{he13}.

\section*{Acknowledgments}
The authors thank Y. Tanaka and K. T. Law for fruitful discussion.
This work is supported by the ``Topological Quantum Phenomena" (No. 22103005) Grant-in Aid for Scientific Research on Innovative Areas from the Ministry of Education, Culture, Sports, Science and Technology (MEXT) of Japan. 
M.S. is supported by Grant-in-Aid for Scientific Research B (No. 25287085) from Japan Society for the Promotion of Science (JSPS).

\appendix
\section{General formula of charge conductance for the effective model}
\label{derivation}

T-matrix Eq. (\ref{tmat}) is given by
\begin{align}
T_{\sigma \mathrm e, \sigma \mathrm h}(\omega)
 &=\left[
 -2 ( \tilde t_{\sigma 1}^2 + \tilde t_{\sigma 2}^2) \omega 
-i 4 ( \tilde t_{\sigma 1}^2 \Gamma_{22} + \tilde t_{\sigma 2}^2 \Gamma_{11}) \right.
\nonumber\\ & \left.\quad
+ i 4 \tilde t_{\sigma 1} \tilde t_{\sigma 2} (\Gamma_{12} +
 \Gamma_{21})\right]/X,
\\
T_{\sigma \mathrm e, -\sigma \mathrm h}(\omega)
&=\left[
-2 (\tilde t_{\sigma 1} \tilde t_{-\sigma 1} 
+ \tilde t_{\sigma 2} \tilde t_{-\sigma 2} ) \omega\right.
\nonumber\\ & \quad
-i 4 (\tilde t_{\sigma 1} \tilde t_{-\sigma 1} \Gamma_{22} + \tilde t_{\sigma 2} \tilde t_{-\sigma 2} \Gamma_{11})
\nonumber\\ & \quad
+ i 2 (\tilde t_{\sigma 1} \tilde t_{-\sigma 2} + \tilde t_{\sigma 2} \tilde t_{-\sigma 1}) (\Gamma_{12} + \Gamma_{21})
\nonumber\\ & \left.\quad
-i 2 (\tilde t_{\sigma 1} \tilde t_{-\sigma 2} - \tilde t_{\sigma 2} \tilde t_{-\sigma 1}) E_{\rm M}\right]/X,
\end{align}
where
the denominator $X$ of the T--matrix is given by
\begin{align}
X = (\omega + i 2 \Gamma_{11}) (\omega + i 2 \Gamma_{22}) +(\Gamma_{11}+\Gamma_{22})^2 - E_{\rm M}^2,
\end{align}
with $\Gamma_{ij} = \sum_\sigma \tilde t^*_{\sigma i} \tilde t_{\sigma j}$.

\section{Tunneling conductance of the NS junction}

In this Appendix, we explain how to calculate the tunneling conductance
for a NS junction of the superconducting QAH insulator.
N (S) is located in $x<0$ ($x>0$).
The Hamiltonian of the NS junction is given by Eqs. (\ref{ns1}) and (\ref{ns2}).
The wave functions $\psi$ of the scattering state for $x<0$ and $x>0$
has the following form,
\begin{align}
 \psi_{\sigma} (x < 0) &= \chi_{\sigma \mathrm e} e^{i k_{\rm e} x }
 + \sum_{\sigma' \tau'}r_{\sigma \sigma' \tau'} \chi_{\sigma' \tau'} e^{-i k_{\tau'} x},
 \\
 \psi_{\sigma} (x>0)
 &=
 \sum_{\mu} t_{\sigma \mu} \bm u_{{\mu}}(q_{\mu}) e^{i q_{\mu} x},
\end{align}
where $\sigma$ is spin of the incident electron, 
$k_{\tau} = \tau \sqrt{2m_{\rm N} (\mu_{\rm N} + \tau E)}$, 
$\chi_{\sigma' \mathrm e}$ and $\chi_{\sigma' \mathrm h}$ are eigenvectors for electron and hole states in N with spin $\sigma'$, and $\tau = \mathrm e = +1$, $\tau = \mathrm h = -1$.
$\bm u_{\mu}(q_\mu)$ is the eigenvector of $H_{\rm SQAH}$:
\begin{align}
 H_{\rm SQAH}(q_{\mu}) \bm u_{\mu}(q_{\mu})
 = E_{\mu}(q_{\mu}) \bm u_{\mu}(q_{\mu}).
\end{align}
The energy satisfies
$E_{\mu}(q_\mu)=E$, where $\mu$ is the band index.
The momentum is determined by
$\mathrm{Im} (q_\mu) > 0$ for an evanescent state,  $\partial E_\mu(q_\mu)/\partial q_\mu > 0$ and $\mathrm{Im} (q_\mu) = 0$ for a propagating state. 
Note that there are four states satisfying the above conditions.

The tunneling conductance is given by
\begin{align}
 G = \frac{e^2}{h}
 \left[
  2 - \sum_{\sigma\sigma'} 
\left(
|r_{\sigma \sigma' \mathrm e}|^2
  -
  \left|\frac{k_{\rm h}}{k_{\rm e}}
  \right|
|r_{\sigma \sigma' \mathrm h}|^2
  \right)
 \right].
\end{align}
The reflection coefficient $r_{\sigma \sigma' \tau}$ is deduced by applying the continuity condition on the wave function:
\begin{align}
 \psi_\sigma(-0) &= \psi_\sigma(+0),
 \\
 v_{\rm N} \psi_\sigma(-0) &= v_{\rm SQAH} \psi_\sigma(+0),
\end{align}
with 
\begin{align}
v_{\rm N} &= \frac{\partial H(x < 0)}{\partial (-i \partial_x)}
\nonumber\\
&= \frac{-i \partial_x}{m_{\rm N}},
\\
v_{\rm SQAH} &= 
	\frac{\partial H(x>0)} {\partial (-i \partial_x)} 
\nonumber\\
&= 2B (-i \partial_x) \sigma_z \tau_z + A \sigma_x,
\end{align}
being the velocity operators in N and S, respectively.

\bibliographystyle{model1a-num-names}
\bibliography{majorana_interference}

\clearpage

\end{document}